\documentclass[aps, 10pt, prl,notitlepage, twocolumn, superscriptaddress,nofootinbib]{revtex4-2}

\usepackage{amsmath}
\usepackage{amssymb}
\usepackage{amsfonts}
\usepackage[utf8]{inputenc}
\usepackage[T1]{fontenc}
\usepackage{mathrsfs}
\usepackage{hyperref}
\usepackage{anyfontsize}

\usepackage[usenames,dvipsnames]{xcolor}

\usepackage{newtxtext}
\usepackage{newtxmath}
\usepackage{tensor}

\usepackage{graphicx}
\usepackage{epsfig}
\usepackage{epstopdf}
\usepackage{lipsum}

\usepackage{natbib}

\usepackage[normalem]{ulem}

\def \nn  {\nonumber}

\def\jnl@style{\it}
\def\aaref@jnl#1{{\jnl@style#1}}

\def\aaref@jnl#1{{\jnl@style#1}}

\def\aj{\aaref@jnl{AJ}}                   
\def\apj{\aaref@jnl{ApJ}}                 
\def\apjl{\aaref@jnl{ApJ}}                
\def\apjs{\aaref@jnl{ApJS}}               
\def\apss{\aaref@jnl{Ap\&SS}}             
\def\aap{\aaref@jnl{A\&A}}                
\def\aapr{\aaref@jnl{A\&A~Rev.}}          
\def\aaps{\aaref@jnl{A\&AS}}              
\def\mnras{\aaref@jnl{Mon.~Not.~Roy.~Astron.~Soc.}}             
\def\prd{\aaref@jnl{Phys.~Rev.~D}}        
\def\prc{\aaref@jnl{Phys.~Rev.~C}}  
\def\prl{\aaref@jnl{Phys.~Rev.~Lett.}}    
\def\qjras{\aaref@jnl{QJRAS}}             
\def\skytel{\aaref@jnl{S\&T}}             
\def\ssr{\aaref@jnl{Space~Sci.~Rev.}}     
\def\zap{\aaref@jnl{ZAp}}                 
\def\nat{\aaref@jnl{Nature}}              
\def\aplett{\aaref@jnl{Astrophys.~Lett.}} 
\def\apspr{\aaref@jnl{Astrophys.~Space~Phys.~Res.}} 
\def\physrep{\aaref@jnl{Phys.~Rep.}}      
\def\physscr{\aaref@jnl{Phys.~Scr}}       
\def\commat{\aaref@jnl{Comm.~Math.~Phys.}}              
\def\science{\aaref@jnl{Science}}               
\def\cqg{\aaref@jnl{Classical Quant.~Grav.}}            
\def\jpcs{\aaref@jnl{JPCS}}                                     
\def\ijmpd{\aaref@jnl{Int.~J.~Mod.~Phys.~D}}                    
\def\grg{\aaref@jnl{Gen.~Relat.~Gravit.}}               
\def\rpp{\aaref@jnl{Rep.~Prog.~Phys.}}          
\def\npa{\aaref@jnl{Nucl.~Phys.~A}}        
\def\lrr{\aaref@jnl{Living Rev.~Rel.}}                   
\def\jcap{\aaref@jnl{J.~Cosmology Astropart.~Phys.}}    
\def\rmp{\aaref@jnl{Rev.~Mod.~Phys.}}   

\allowdisplaybreaks[1]

\begin{document}

	\title{Dynamical Formation of Scalarized Black Holes and Neutron Stars through Stellar Core Collapse}
	
	\author{Hao-Jui Kuan}
	\email{hao-jui.kuan@uni-tuebingen.de}
	\affiliation{Theoretical Astrophysics, Eberhard Karls University of T\"ubingen, T\"ubingen 72076, Germany}
	\affiliation{Department of Physics, National Tsing Hua University, Hsinchu 300, Taiwan}
	
	\author{Daniela D. Doneva}
	\email{daniela.doneva@uni-tuebingen.de}
	\affiliation{Theoretical Astrophysics, Eberhard Karls University of T\"ubingen, T\"ubingen 72076, Germany}

	\author{Stoytcho S. Yazadjiev}
	\email{yazad@phys.uni-sofia.bg}
	\affiliation{Theoretical Astrophysics, Eberhard Karls University of T\"ubingen, T\"ubingen 72076, Germany}
	\affiliation{Department of Theoretical Physics, Faculty of Physics, Sofia University, Sofia 1164, Bulgaria}
	\affiliation{Institute of Mathematics and Informatics, 	Bulgarian Academy of Sciences, 	Acad. G. Bonchev Street 8, Sofia 1113, Bulgaria}

	\begin{abstract}
		In a certain class of scalar-Gauss-Bonnet gravity, the black holes and the neutron stars can undergo spontaneous scalarization -- a strong gravity phase transition triggered by a tachyonic instability due to the nonminimal coupling between the scalar field and the spacetime curvature. Studies of this phenomenon have, so far, been restricted mainly to the study of the tachyonic instability and stationary scalarized black holes and neutron stars. To date, no realistic physical mechanism for the formation of isolated scalarized black holes and neutron stars has been proposed. We study, for the first time, the spherically symmetric fully nonlinear stellar core collapse to a black hole and a neutron star in scalar-Gauss-Bonnet theories allowing for a spontaneous scalarization. We show that the core collapse can produce scalarized black holes and scalarized neutron stars starting with a nonscalarized progenitor star. The possible paths to reach the end (non)scalarized state are quite rich leading to interesting possibilities for observational manifestations.
	\end{abstract}
	
	\maketitle
	
	\emph{Introduction.---} 
	Among the modified gravity theories, the scalar-Gauss-Bonnet (SGB) gravity takes a special place. This theory is an extension of general relativity (GR) and contains a dynamical scalar field coupled to the Gauss-Bonnet (GB) invariant. The roots of SGB gravity lay in the low-energy limit of quantum gravity and unification theories \cite{Zwiebach:1985uq,Gross:1986mw} as well as in the effective field theories. As in GR, the field equations of SGB gravity are of second order and the theory is free from ghosts. There exists a particular class of SGB theories which gives rise to spontaneously scalarized black holes (BHs) and neutron stars (NSs) \cite{Doneva_2018,Silva_2018,Antoniou:2017acq,Doneva_2018a}. More precisely, the spacetime curvature itself can induce a tachyonic instability that spontaneously scalarizes the black holes or the neutron stars. This class of SGB theories is indistinguishable from GR in the weak field limit and is yet unconstrained by gravitational wave (GW) observations. Since this interesting phenomenon is the only known dynamical mechanism for endowing black holes and neutron stars with scalar hair, it has attracted a lot of interest in recent years (though a novel nonlinear mechanism that can lead to dynamical formation of scalar hair beyond the standard spontaneous scalarization was recently proposed by \cite{Doneva21}).
	
	Thanks to the efforts of many researchers, the spontaneous curvature induced scalarization in SGB gravity has been extensively studied, and now, we have a pretty good understanding of this phenomenon. In particular, the tachyonic instability that triggers the spontaneous scalarization is, to a large extent, well understood \cite{Doneva_2018,Silva_2018,Antoniou:2017acq,Doneva_2018a,Dima:2020yac,Doneva:2020nbb,Doneva_2020a}. The same applies to the static or stationary BH and NS solutions that are the end states of the tachyonic instability \cite{Doneva_2018,Silva_2018, Doneva_2018a,Cunha_2019,Collodel:2019kkx,Herdeiro:2020wei,Berti:2020kgk}. Even the highly nonlinear dynamics of the curvature induced spontaneous scalarization is, to some extent, well understood from a mathematical point of view \cite{Ripley_2020,Doneva:2021dqn} including the dynamical descalarization during black hole merger \cite{Silva:2020omi,East:2021bqk} (see, also, \cite{Benkel:2016kcq}). However, there is a very important link missing in our understanding of the curvature induced spontaneous scalarization. Up to now, no realistic physical scenario for the formation of isolated, scalarized BHs and NSs has been investigated. The purpose of the present Letter is to show that the scalarized compact objects can be formed under gravitational core collapse (CC) of a nonscalarized progenitor star and to explore the different scenarios depending on the theory parameters and the progenitors.
	
	Numerical simulations that demonstrate the core-collapse process with a scalarized compact object as the remnant have been limited in scalar-tensor theories\cite{Novak:1999jg,Gerosa:2016fri,Sperhake:2017itk,Cheong:2018gzn,Rosca-Mead:2019seq,Geng:2020slq,Rosca-Mead:2020ehn} until now. In this case, though, scalarization of BHs is typically not possible (for interesting, though not so astrophysically relevant exceptions, we refer the reader to \cite{Stefanov:2007eq,Doneva:2010ke,Cardoso:2013fwa}). Since SGB theories allow for nonlinear development of BH scalar hair, they provide a richer phenomenology of the core collapse. However, the complexity of the field equations is significantly increased. In the present Letter, we go beyond the commonly employed decoupling limit approximation \cite{Benkel:2016kcq,Doneva:2021dqn,Silva:2020omi} and consider the coupled evolution of the spacetime, matter and the scalar field. 
	
	\emph{Gauss-Bonnet theory.---}
	The action of the sGB gravity in the presence of  matter is the following:  
	\begin{eqnarray}
		S=&&\frac{1}{16\pi}\int d^4x \sqrt{-g} 
		\Big[R - 2\nabla_\mu \varphi \nabla^\mu \varphi  + \\
		&& \hspace{2cm} + \lambda^2 f(\varphi){\cal R}^2_{\text{GB}} \Big] + S_{\rm matter}  (g_{\mu\nu},\Psi_m) .\label{eq:quadratic} \notag
	\end{eqnarray}
	where $f(\varphi)$ is the coupling function of the scalar field $\varphi$ to the GB invariant ${\cal R}^2_{\text{GB}}=R^2 - 4 R_{\mu\nu} R^{\mu\nu} + R_{\mu\nu\alpha\beta}R^{\mu\nu\alpha\beta}$. The GB coupling constant $\lambda$ has  dimension of length, and the matter fields are collectively denoted by $\Psi_m$.

	The field equations derived from the action are 
	\begin{eqnarray}\label{FE}
		&&R_{\mu\nu}- \frac{1}{2}R g_{\mu\nu} + \Gamma_{\mu\nu}= 2\nabla_\mu\varphi\nabla_\nu\varphi -  g_{\mu\nu} \nabla_\alpha\varphi \nabla^\alpha\varphi  + 8\pi T_{\mu\nu},\nonumber\\
		\\
		&&\nabla_\alpha\nabla^\alpha\varphi= -  \frac{\lambda^2}{4} \frac{df(\varphi)}{d\varphi} {\cal R}^2_{\text{GB}},\label{SFE}
	\end{eqnarray}
	where  $T_{\mu\nu}$ is the matter energy-momentum tensor that can be proven to satisfy $\nabla^{\mu}T_{\mu\nu}=0$ and
	\begin{eqnarray}\label{eq:gamdef}
		\Gamma_{\mu\nu}= &&- R(\nabla_\mu\Psi_{\nu} + \nabla_\nu\Psi_{\mu} ) - 4\nabla^\alpha\Psi_{\alpha}\left(R_{\mu\nu} - \frac{1}{2}R g_{\mu\nu}\right) \nonumber  \\
		&& + 4R_{\mu\alpha}\nabla^\alpha\Psi_{\nu} + 4R_{\nu\alpha}\nabla^\alpha\Psi_{\mu}  \nonumber\\
		&&- 4 g_{\mu\nu} R^{\alpha\beta}\nabla_\alpha\Psi_{\beta} 
		+ \,  4 R^{\beta}_{\;\mu\alpha\nu}\nabla^\alpha\Psi_{\beta}
	\end{eqnarray}  
	with $\Psi_{\mu}= \lambda^2 [df(\varphi)/d\varphi]\nabla_\mu\varphi$ .

	We consider asymptotically flat spacetimes with zero cosmological value of the scalar field $\varphi_{\infty}=0$. The  GB coupling function $f(\varphi)$ allowing for spontaneous scalarization has to obey the condition  $(df/d\varphi)(0)=0$.
	In addition we can impose $f(0)=0$ and $(d^2f/d\varphi^2)(0)=\epsilon$ with $\epsilon=\pm 1$.
	
	\emph{Evolution equations.---}
	In the present letter, we will study the coupled evolution of the scalar field, the matter, and the spacetime in spherical symmetry, that is a simplification also adopted in previous studies on core collapse in alternative theories of gravity \cite{Novak:1999jg,Gerosa:2016fri,Sperhake:2017itk,Cheong:2018gzn,Rosca-Mead:2019seq,Geng:2020slq,Rosca-Mead:2020ehn}. The resulting equations are quite lengthy, and they are given in the Appendix. For the numerical calculations, we employ the \texttt{GR1D} code \cite{OConnor:2009iuz,Gerosa:2016fri} and implement a significant modification in order to deal with the SGB field equations in the fashion of \cite{Ripley_2020,Ripley19c} (see the Appendix for details).
	
	Even though different coupling functions satisfying the scalarization criteria can be introduced, the disparity between the stable scalarized solutions (if exist that is not the case for example for the simplest choice $f(\varphi)=\varphi^2$) and their dynamics for different couplings is mainly quantitative \cite{Doneva:2018rou,Doneva:2021dqn}. From a numerical point of view, one of the least problematic, generic, and widely used coupling functions for which stable scalarized BHs can exist is the following \cite{Doneva_2018,Doneva_2018a,Cunha_2019,Herdeiro:2020wei,East:2021bqk}
	\begin{eqnarray}\label{eq:coupling}
		f(\varphi)=  \frac{\epsilon}{2\beta} \left[1-\exp(-\beta\varphi^2)\right], \,\,\, \epsilon=\pm 1,
	\end{eqnarray}	
	where $\beta>0$ is a parameter that has not been constrained by the observations yet. Based on \cite{Doneva:2021dqn,Doneva:2018rou}, we expect that the main results in this Letter will remain qualitatively similar for a broad class of couplings that lead to scalarization.
	
	\emph{Possible core-collapse scenarios.---}
	In the simulations, a hybrid equation of state (EOS) that splits the pressure and density into the cold and the thermal parts, viz.~$p=p_c + p_{\rm th}$ and $e=e_c + e_{\rm th}$, is used. Here, $e$ is the specific internal fluid energy. A piecewise polytrope is employed for the cold part EOS which consists of two polytropes with indexes $\Gamma_1$ and $\Gamma_2$, stitched at nuclear density $\rho_{\rm nucl}=2\times 10^{14} g\, cm{^{-3}}$ \cite{OConnor:2009iuz}. For the thermal part, we assume $p_{\rm th}=(\Gamma_{\rm th} -1)\rho (e-e_c)$. As initial data for the scalar field, we consider a Gaussian pulse with a mean of $200$ km, minute amplitude of $10^{-10}$, and standard deviation of $100$ km, while we note that the results are independent of the explicit form of the initial data.
	
	\begin{figure}
		\centering
		\includegraphics[scale=0.7]{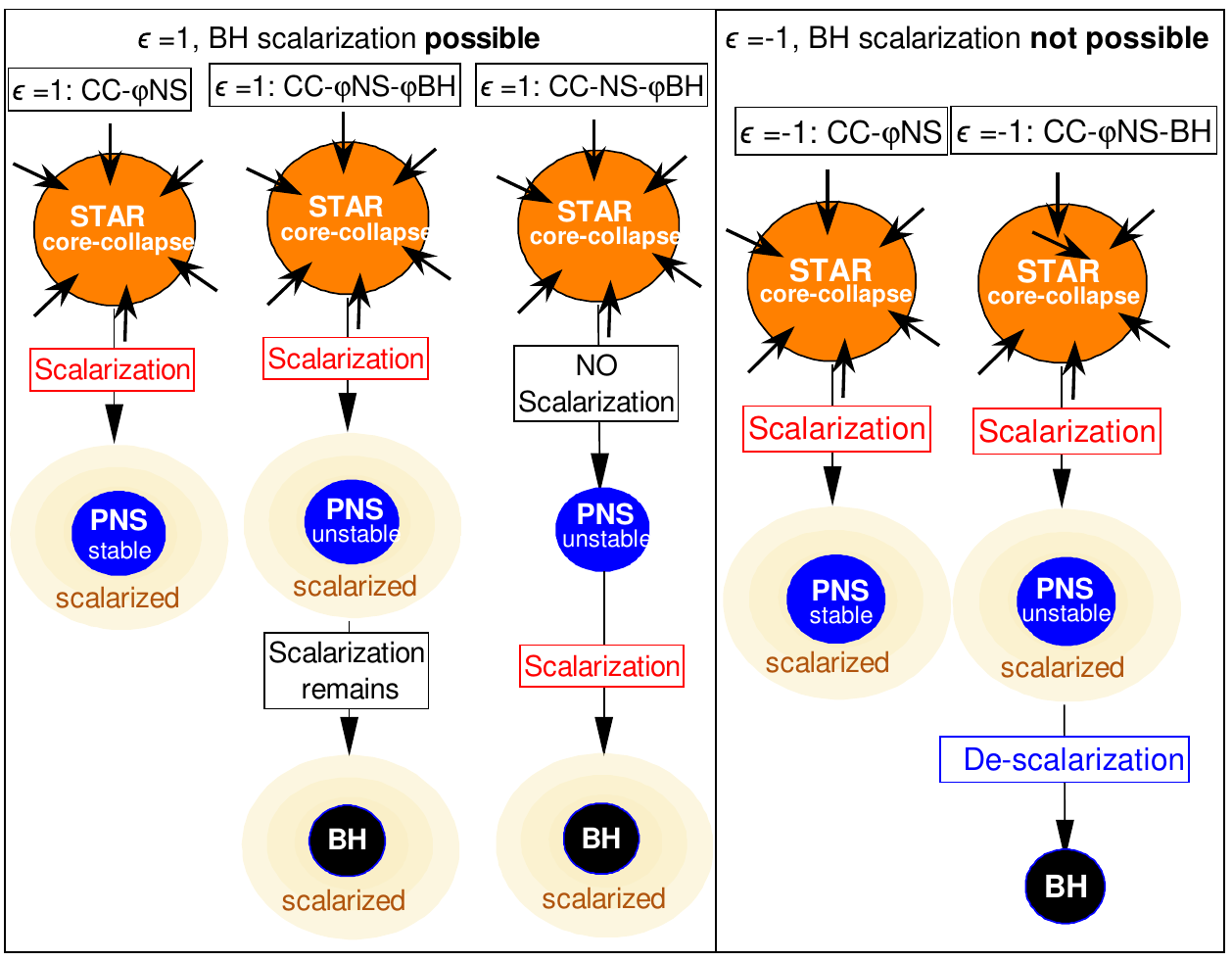}
		\caption{Possible outcomes of stellar core collapse in SGB gravity. }
		\label{fig:CC_Outcome}	
	\end{figure}
	
	Depending on the progenitors, the coupling constants $\beta$ and $\lambda$, and the sign of $\epsilon$, the final outcome of the core collapse and the path to reach it can vary significantly. Figure \ref{fig:CC_Outcome} represents all possible outcomes, which are divided into two major classes: $\epsilon=1$ and $\epsilon=-1$. While in the former case ($\epsilon=1$), both NSs and BHs can scalarize, in the latter case ($\epsilon=-1$) scalarization is not possible for static BHs and only NSs can develop nontrivial scalar field \cite{Doneva_2018,Doneva_2018a,Silva_2018}. In addition, scalar field can develop during the formation of either the protoneutron star (PNS) or the BH. For $\epsilon=-1$, scalarization can appear only temporarily at the PNS stage.
	Let us note that if a relatively fast rotating BH is formed after the core collapse, it might be possible to have another scenarios for  $\epsilon=-1$ because of the spin-induced scalarization \cite{Dima:2020yac,Doneva:2020nbb,Doneva_2020a,Herdeiro:2020wei,Berti:2020kgk}; however, production of a rapidly rotating protoneutron star after a core collapse seems to be a rare event \cite{Popov:2012ng,Noutsos:2013ce}.
	
	We set $\lambda$ and $\beta$ to appropriate values for the aforementioned scenarios to be realized. The critical central energy density $\rho_{\rm bifurc}$, above which scalarization is possible, is controlled by $\lambda$, while $\beta$ is responsible for the ``degree'' of scalarization \cite{Doneva_2018a}. In general, larger $\beta$ leads to weaker scalar fields (for a fixed $\lambda$).
	
	\begin{figure}
		\centering
		\includegraphics[scale=0.46]{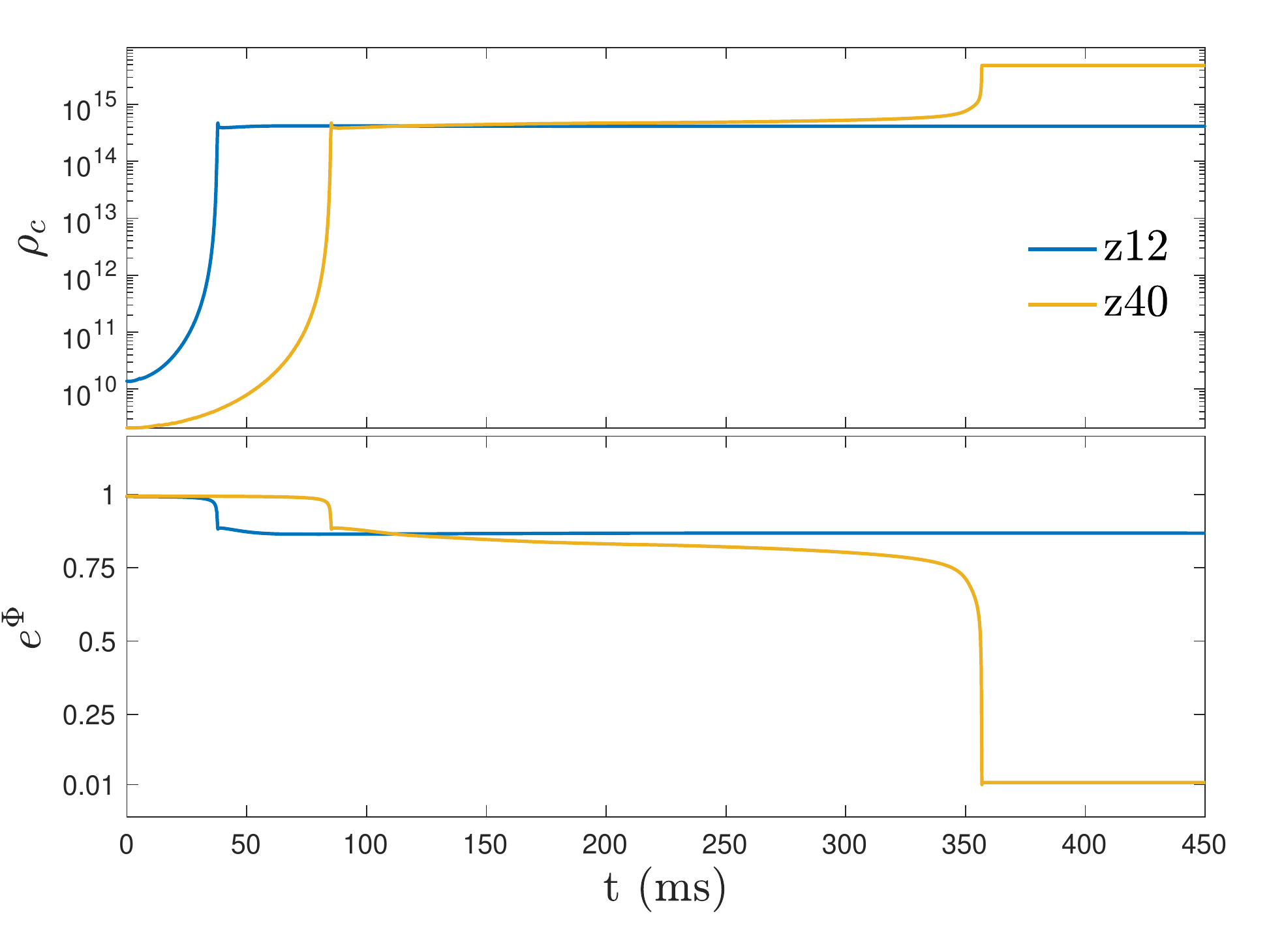}
		\caption{Central densities (\text{top panel}) and central values of redshift (\text{bottom panel}) of supernova progenitors $z12$ (blue line) and $z40$ (yellow line) \cite{Woosley:2007as} as functions of time in GR. 
			The formation of a black hole is numerically defined to form as central redshift $e^{\Phi}<0.01$.}
		\label{fig:evo}	
	\end{figure}
	
	\emph{The progenitors.---} 
	For the progenitors of stellar core collapse, we use some models provided in Woosley \& Heger's catalog \cite{Woosley:2007as}. The simulated results are qualitatively the same. In particular, for the remnant with the central density larger than the $\lambda$-dependent $\rho_{\rm bifurc}$, scalarization is observed, whose degree then depends on $\beta$. Without loss of generality, here, we present the simulations with two progenitors $z12$ and $z40$, having primordial metallicity, that have also been investigated in scalar-tensor theory \cite{Gerosa:2016fri,Sperhake:2017itk,Cheong:2018gzn,Geng:2020slq} and are good for comparison.
	
	The model $z12$ has the zero-age-main-sequence (ZAMS) mass of $M=12M_\odot$ and collapses to a stable PNS. The steep density gradient outside its iron core results in a low accretion rate after bounce. On the other hand, the model $z40$, more massive with $M=40M_\odot$, evolves into a short-lived PNS which then collapses to a BH as its shallow density leads to high accretion rate after bounce. Simulations adopting an approximation of deleptonization show the index for the softer, cold piece of EOS to be $\Gamma_1\sim1.3$ \cite{Dimmelmeier07}, while  $\Gamma_2=2.5-3$ is found to be approximants for some realisitc finite-temperature EOS \cite{Dimmelmeier08}, e.g., Shen \cite{Shen11} and Lattimer and Swesty \cite{LS} EOS, for the stiffer, cold component. The thermal description of a mixture of relativistic and non-relativistic gas can be translated to $4/3<\Gamma_{\text{th}}<5/3$. We consider, as canonical values, $\Gamma_1=1.3$, $\Gamma_2=2.3$, and $\Gamma_{\text{th}}=1.35$  following, e.g., \cite{Gerosa:2016fri,Sperhake:2017itk,Cheong:2018gzn,Rosca-Mead:2019seq,Geng:2020slq,Rosca-Mead:2020ehn}. We note that our simulations show that different combinations of parameters, each falling in the aforementioned range, does not phenomenologically alter the scalar field dynamics. The evolution of the progenitors’ central density and redshift in GR is shown in Fig.~\ref{fig:evo}. The evolution will remain qualitatively and quantitatively similar for weak to moderate scalar fields. A significant change is expected for strong scalar fields, but they might lead to loss of hyperbolicity and deserve special attention (see below).
	
	\begin{figure}
		\centering
		\includegraphics[scale=0.46]{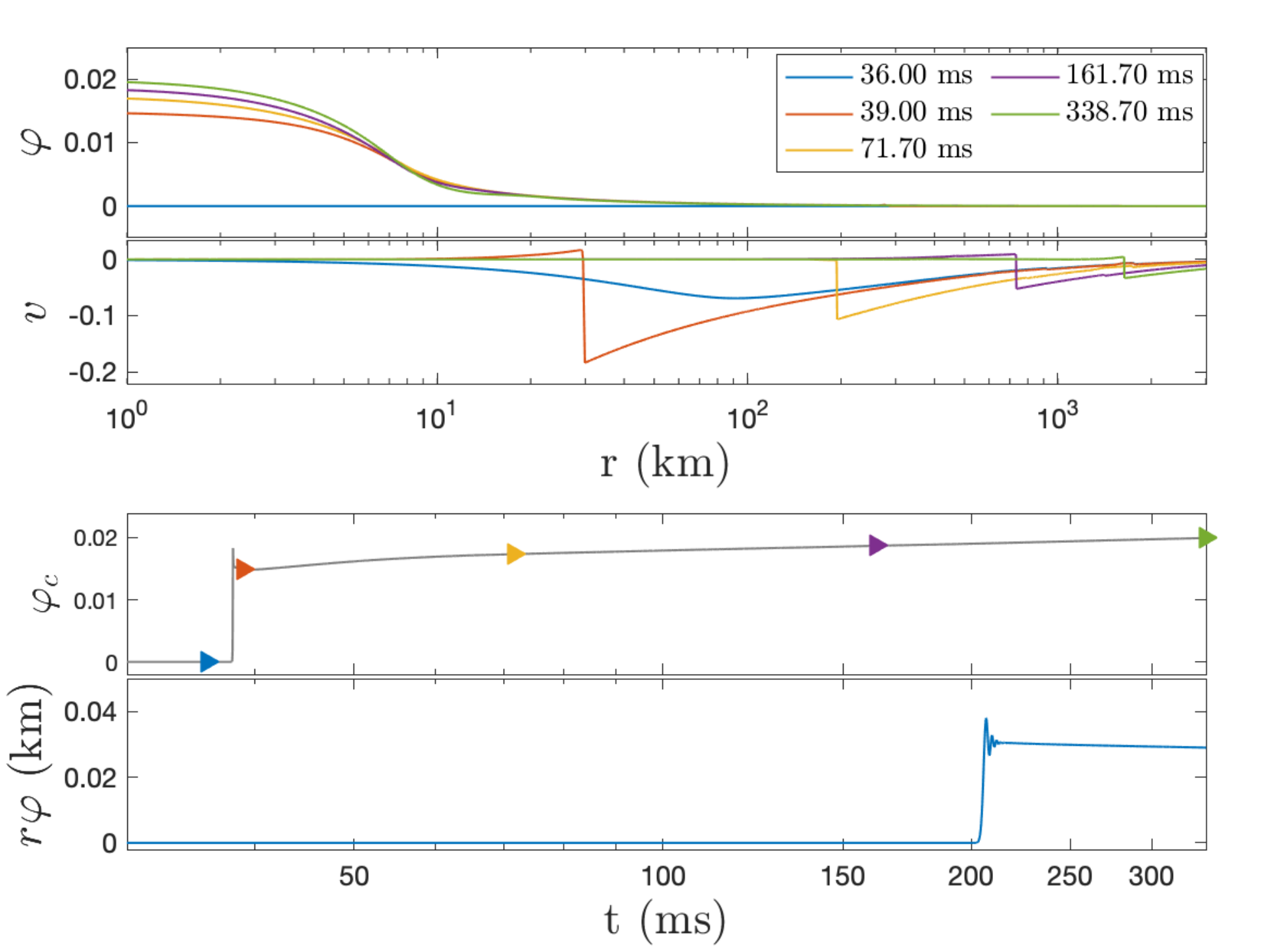}
		\caption{(scenario ``$\epsilon=-1$:~CC-$\varphi$NS'') Temporal snapshots of scalar field $\varphi$ and fluid velocity $v$ as functions of the distance from the core $r=0$ for the $z12$ progenitor are plotted in the upper panel. In the lower panel, the evolution of the central value of the scalar field $\varphi_c$ and the scalar charge $r\varphi$ taken at a very large distance, $50$ $000$ km, are displayed. Markers in the bottom panels indicate the time of snapshots having the same colors in the corresponding upper panels. We have taken $\epsilon=-1, \lambda=80$, and $\beta=7000$.}
		\label{fig:NS_remnant1}
	\end{figure}

	\emph{Core collapse to a scalarized neutron star.---}	
	For the model $z12$, the remnant of the collapse is PNS, which may be imbued with scalar field for both $\epsilon=1$ and $\epsilon=-1$. However, the profiles of the scalar hair differ considerably from each other. Here, we examine the case with $\epsilon=-1$ while $\epsilon=1$ is discussed in the Appendix for completeness.
	Fig.~\ref{fig:NS_remnant1} shows a spontaneous scalarization upon the formation of the PNS remnant. We see that the PNS forms at $\sim 38$ ms resulting in a rapid development of a scalar field. Afterward, the hot mantle ranging from $\sim 10-20$ km still accretes matter onto the remnant as illustrated by the velocity profile; the extent to which the scalar field is excited is enhanced with increasing compactness [the third from top panel of Fig.~\ref{fig:NS_remnant1}]. Clearly, the scalarization first develops in the immediate vicinity of the remnant, and then, it propagates to infinity as the profile of $\varphi$ settles to a quasiequilibrium one. The spontaneous appearance of $r\varphi$ at the associated retarded time indicates that the scalar wave propagates at the speed of light in SGB. We have checked $r\varphi$ at three different places, viz.~$500$ km, $5000$ km, and $50$ $000$ km, to confirm it does saturate. 
	
	As it is well known, in certain regions of the parameter space, the SGB field equations lose their hyperbolic character \cite{Ripley19a,Ripley_2020,East:2021bqk}. How to deal with this problem in SGB theories is still an open question and is beyond the scope of the present Letter. In our simulation, we observed that, if we fixed $\lambda$, there is a threshold $\beta$ below which the system loses its hyperbolic character. In practice, this effectively limits the strength of the scalar field since larger $\beta$ leads to weaker scalar field (for fixed $\lambda$). The loss of hyperbolicity and the region where this happens is discussed further in the Appendix. 
	
	The parameter $\beta$ for the simulations presented in this Letter, including Fig.~\ref{fig:NS_remnant1}, is chosen to be close to this threshold $\beta$, and the results remain qualitatively similar for larger $\beta$. For the presented model, the change of the metric function with respect to GR is relatively small, of the order of 1\%, and the total energy of the scalar field reach $\lesssim 1\%$ of the compact object mass.
	
	The threshold for loss of hyperbolicity will also depend on the progenitor properties, the employed EOS, and the parameters of the theory.
	 A thorough investigation of this threshold as different parameters of the system are varied will be considered in a subsequent paper.

	\emph{Collapse to a black hole.---}
	Now, we turn to study collapses with a middle PNS stage followed by a BH formation due to continuous accretion by exhibiting the evolution of $z40$. As plotted in Fig.~\ref{fig:evo}, the PNS forms at $\sim 85$ ms  and lives for $\sim 300$ ms. Afterward, the BH remnant appears at $\sim 360$ ms, and matter and scalar field cease evolving inside the event horizon once it has formed. The parameter $\beta$ is chosen to be close to the threshold where loss of hyperbolicity is observed (for the corresponding $\lambda$). 
	
\begin{figure}
		\centering
		\includegraphics[scale=0.46]{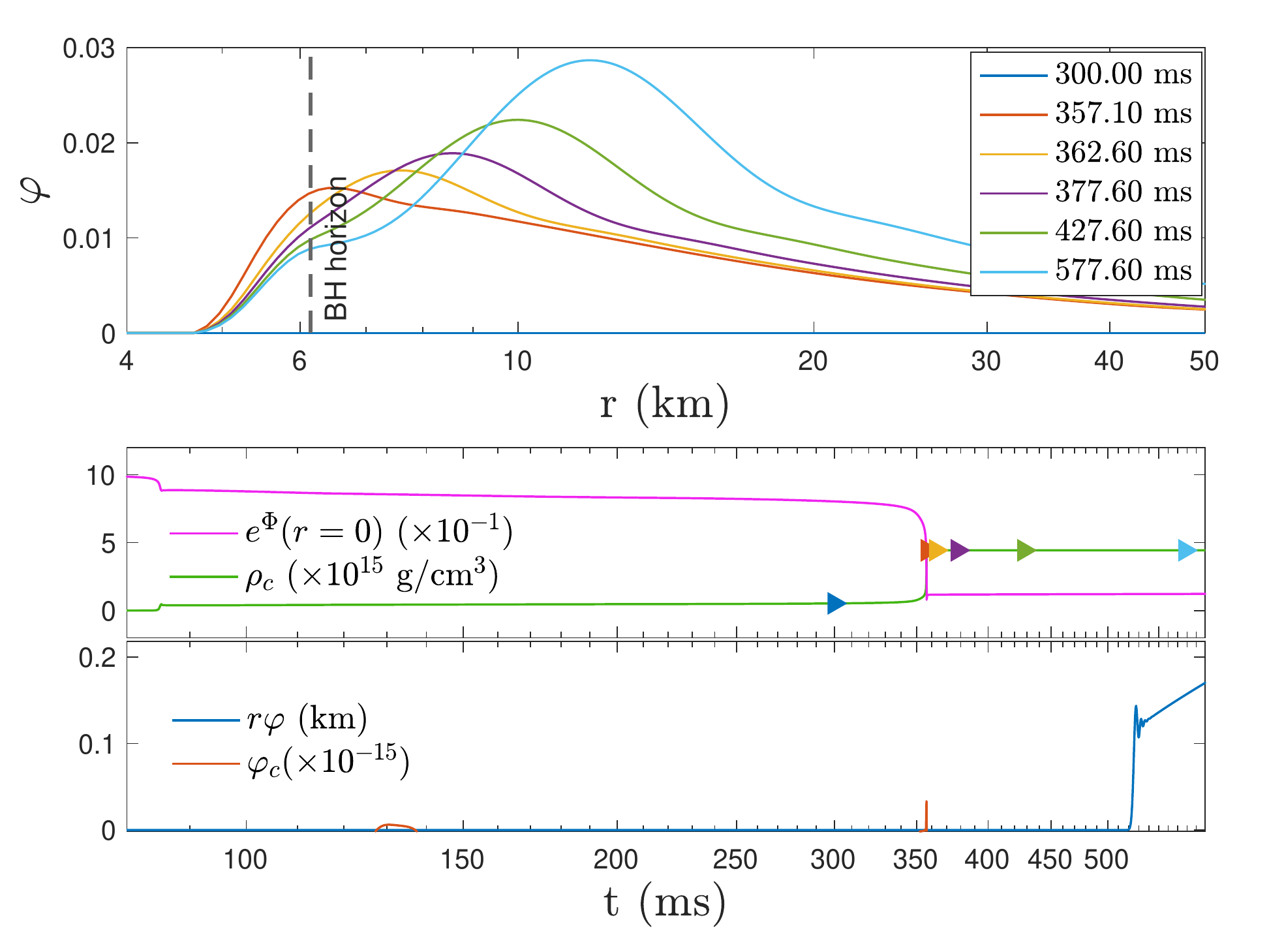}
		\caption{(scenario ``$\epsilon=1$: CC-NS-$\varphi$BH'' ) Temporal snapshots of scalar field $\varphi$ as a function of the distance from the core $r=0$ is potted for the $z40$ progenitor in the upper panel. In the lower group of panels, central value of scalar field $\varphi_c$ and the scalar charge $r\varphi$ are displayed as functions of time. The notations are the same as in Fig. \ref{fig:NS_remnant1}. We have taken $\epsilon=1$, $\lambda=30,\beta=20$ $000$. }
		\label{fig:NS_BH_remnant1}	
	\end{figure}

	Two channels are possible in collapses leading to a BH for $\epsilon=1$, namely, scalarization is absent or present during the middle PNS stage of the collapse (see Fig.~\ref{fig:CC_Outcome}). In Figs.~\ref{fig:NS_BH_remnant1} the channel ``$\epsilon=1$: CC-NS-$\varphi$BH'' is presented while the other channel ``$\epsilon=1$: CC-$\varphi$NS-$\varphi$BH'' is discussed in the Appendix for completeness. 
	Temporal snapshots of scalar field $\varphi$ as a function of the distance from the center $r=0$ are plotted in the upper panel. In the lower panel, the evolution of the central value of the scalar field $\varphi_c$ and the scalar charge $r\varphi$ are displayed. The evolution of $\varphi_c$ is just symbolic since it is uncertain how the fields ($\rho$, $p$, $\varphi$, \dots) evolve inside the event horizon, and we plot $\varphi_c$ as constant since we freeze the fields interior to the event horizon similar to \cite{Novak:1999jg,Gerosa:2016fri}.
	
\begin{figure}
		\centering
		\includegraphics[scale=0.46]{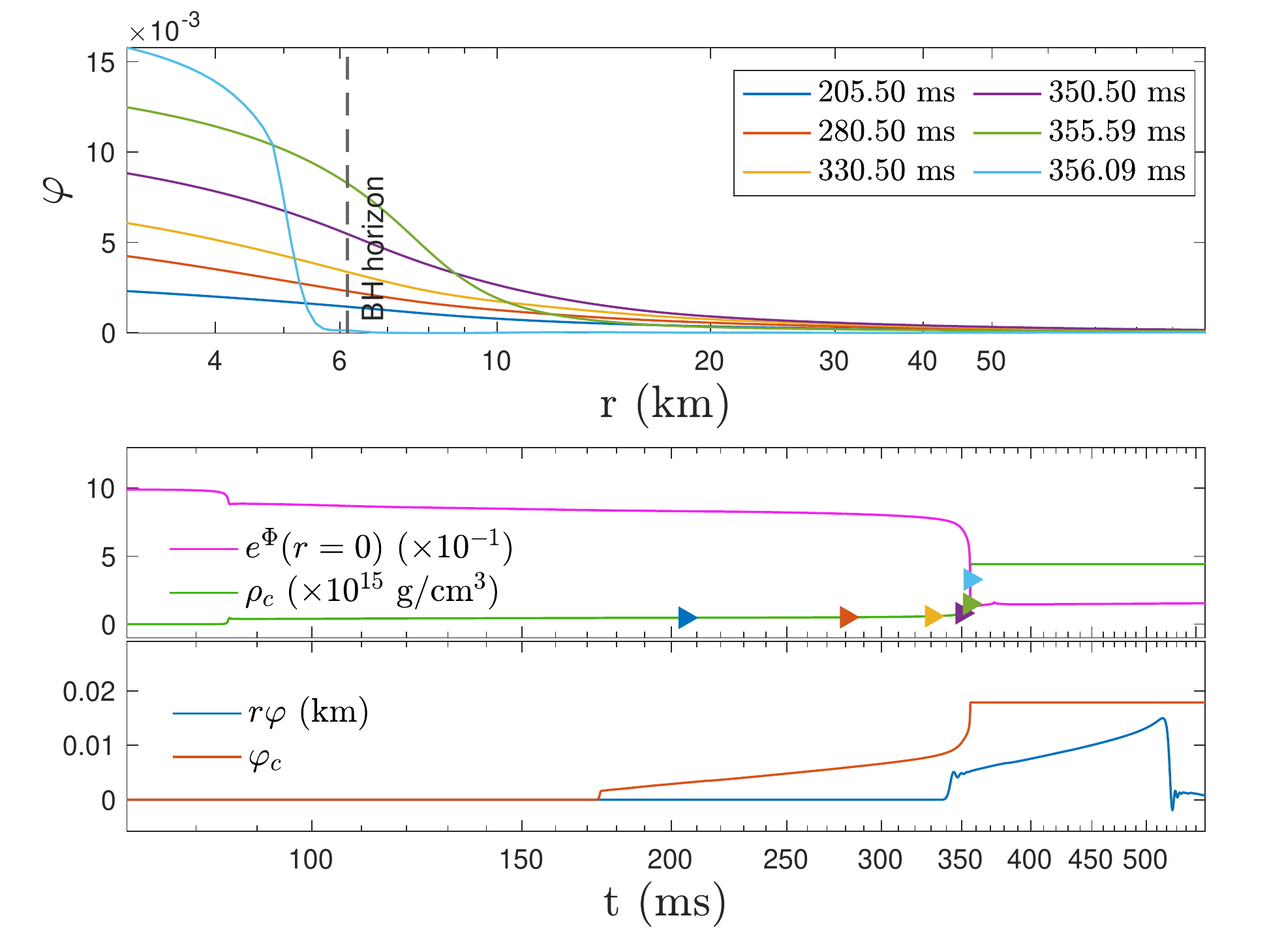}
		\caption{(scenario ``$\epsilon=-1$: CC-$\varphi$NS-BH'') The same as Fig.~\ref{fig:NS_BH_remnant1} but for $\epsilon=-1$, $\lambda=40,\beta=25$ $000$. Scalarization of protoneutron star reveals (light red and yellow curves in the upper panel); nonetheless, once the black hole is formed, the scalar field condensates into the event horizon. Descalarization is apparent as shown by the disappearance of the scalar charge after a period of scalarized protoneutron star stage (blue line in the bottom panel).}
		\label{fig:BH_remnant}	
	\end{figure}

	Scalarization is only possible for NSs for $\epsilon=-1$. Thus, a scalarized PNS will undergo a descalarizaion before it collapses into a BH. This scenario, ``$\epsilon=-1$: CC-$\varphi$NS-BH'', is depicted in Fig.~\ref{fig:BH_remnant}. We see that, upon the formation of the final state BH, the excited scalar field is eliminated (blue curve in the bottom panel of Fig.~\ref{fig:BH_remnant}). During the descalarization, most of the energy stored in the scalar field is swallowed by the BH ($\gtrsim90\%$), while a small portion of the energy is radiated away (curves in the top panel of Fig.~\ref{fig:BH_remnant}). The dynamical result proves that static BHs can not scalarize for $\epsilon<0$ \cite{Doneva_2018,Doneva:2021dqn}, illustrating how the scalar field fades away in SGB.
	
	\begin{figure}
		\centering
		\includegraphics[scale=0.36]{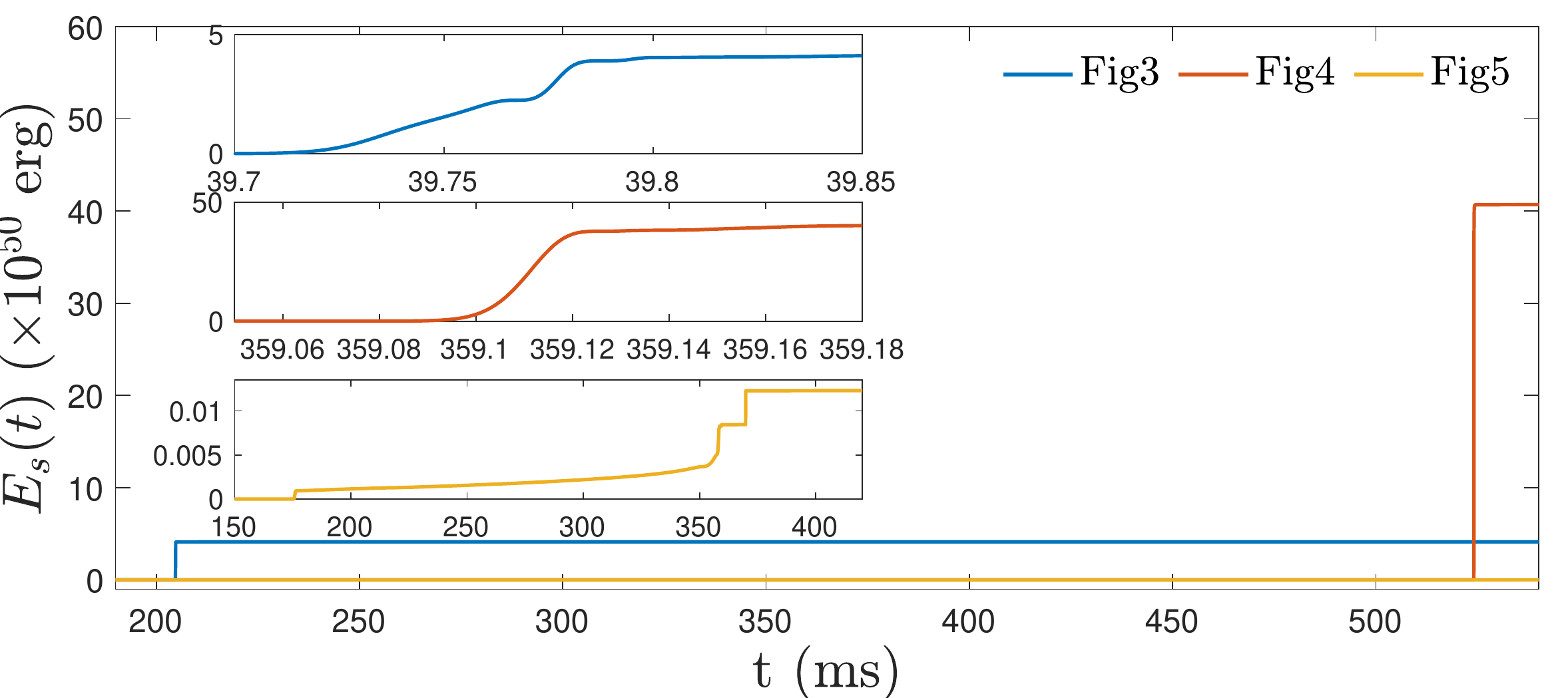}
		\caption{The scalar radiation emitted  during core collapse for the cases presented in Figs. \ref{fig:NS_remnant1}, \ref{fig:NS_BH_remnant1}, and \ref{fig:BH_remnant}. $E_s(t)$ is practically independent of $r$ for large distances, and in the figure, it is extracted at the same point as the scalar charge, i.e., at $50$ $000$ km. The insets represent a magnification around the most interesting regions for each case.}
		\label{fig:scalar_radiation}	
	\end{figure}

	The emitted scalar radiation $E_s$ as a function of time is presented in Fig.~\ref{fig:scalar_radiation}. We see that, at late times, $E_s$ varies in the range $10^{48}-10^{51} {\rm ergs}$ depending on the core collapse model. Given that the typical energy emitted through tensorial GWs during core collapse is $10^{46}-10^{47} {\rm ergs}$ \cite{Radice18,Abdikamalov:2020jzn}, the scalar waves produced during the (de)scalarization of a compact object offer a much more efficient channel of energy loss.
	In the considered SGB gravity, though, the so-called breathing modes, that are potentially detectable, do not exist. If a more general form of sGB gravity is considered such breathing modes can be easily present that would allow us to set constraints on the corresponding SGB theory. Moreover, if one considers the problem beyond the spherical symmetry approximation, other observational manifestations of the scalarization can be expected. This is a very complicated task, though, that has not yet been solved in any alternative theory of gravity.

	\emph{Discussion.---}	
	In this Letter, we presented the first numerical fully nonlinear simulations of the spherically symmetric stellar core collapse to a BH or a NS in SGB theories allowing for a spontaneous scalarization. We showed that, in this process, scalarized BHs and NSs can be produced starting with a nonscalarized progenitor star. In this way, we also demonstrated that the stellar gravitational collapse is the natural physical mechanism for the formation of isolated scalarized compact objects in SGB gravity. There is a variety of collapse scenarios that can be realized with different progenitors, and the SGB parameters that will have different astrophysical manifestations. Thus, with the improvement of the sensitivities of the observations, it might be possible to put strong constraints on the SGB theory. In order to quantify the observational manifestations, though, one has to examine, in much greater detail, the parameters space consisting of the theory parameters, the possible progenitors, and the piecewise polytropic EOS employed in the simulations. This is a study underway.
	
	The consideration of the full system of coupled fluid, metric, and scalar field evolution also allowed us to shed light on the loss of hyperbolicity of the system that is observed for certain ranges of parameters. The system of differential equations turns to a mixed type in the vicinity of the region where shocks appear as the core collapse proceeds. The interpretation of such loss of hyperbolicity is a very involved, open problem in SGB gravity that has not received a proper treatment or interpretation until now. Still, the region where the system behaves well and the Cauchy problem is well defined is large enough, while the scalar field is bounded to relatively low values.

	For the considered SGB theories, in contrast to some standard scalar-tensor theories, for example, the spherically symmetric scalar field dynamics does not lead to the emission of gravitational waves (the so-called breathing modes are absent). Scalar waves will be present, though, and they constitute an additional channel for dissipation of energy and angular momentum that can leave an imprint on the emitted gravitational waves. The collapsing scenarios we considered lead to relatively weak scalar fields, but on the other hand, the scalar radiation is a quite efficient channel of energy dissipation. For the models we considered in the present Letter, the energy dissipation of the scalar waves can be much stronger compared with the expected gravitational wave signal in case of nonspherical core collapse \cite{Radice18,Abdikamalov:2020jzn} that can potentially lead to constraints on Gauss-Bonnet theories.

	\vskip 0.5cm
	D.D.~acknowledges financial support via an Emmy Noether Research Group funded by the German Research Foundation (DFG) under Grant No.~DO 1771/ 1-1. S.~Y.~would like to thank the University of T\"ubingen for the financial support. The partial support by the Bulgarian NSF Grant No.~KP-06-H28/7 and the Networking support by the COST Actions Grants No.~CA16104 and No.~CA16214 are also gratefully acknowledged. H.-J.~K.~acknowledges support from Sandwich Grant (JYP) No.~109-2927-I-007-503 by 351 DAAD and MOST.
	

	\appendix

	\section{Appendix A. Evolution equations for the metric and the fluid variables}\label{sub.evo}
Here we will describe the adopted formalism for evolving in time the spacetime, the fluid and the scalar field. We will follows the notations in \cite{OConnor:2009iuz}.

In a dynamical spherically symmetric spacetimes the metric can be written in the form

\begin{eqnarray}
	ds^2= - e^{2\Phi(r,t)}dt^2 + e^{2\Lambda(r,t)} dr^2 + r^2 (d\theta^2  + \sin^2\theta d\phi^2). 
\end{eqnarray}

It is convenient to introduce the local mass $m(r,t)$ enclosed in a sphere with a radius $r$ at the moment $t$, namely  $m(r,t)=\frac{r}{2}(1-e^{-2\Lambda})$. We will model the matter as a perfect fluid with energy-momentum tensor $T_{\mu\nu}=\rho h u_{\mu}u_{\nu} + pg_{\mu\nu}$ and particle density current $J^\mu= \rho u^{\mu}$. Here $\rho$ is the rest-mass (baryonic) density, $p$ is the fluid  pressure, $h=1 + e + \frac{p}{\rho}$ is the specific enthalpy with $e$ being the specific internal fluid energy, and $u^{\mu}$ is the fluid 4-velocity. Due to the spherical symmetry the 4-velocity of the fluid can be written as
\begin{equation}\label{eq:4vel}
	u^{\mu}= \frac{1}{\sqrt{1-v^2}}(e^{-\Phi}, v e^{-\Lambda},0,0)
\end{equation}   	
with $v=v(t,r)$ being the fluid velocity.	

A well established approach to handle the shock discontinuities appearing in the fluid is the use of the high-resolution shock-capturing schemes. This approach requires the equations of motion to be written in flux conservative form using conserved variables. In our case the conserved variables are $D$, $S^r$ and $\tau$, and they are related to the standard primitive  variables $\rho$, $v$ and $p$ by  
\begin{equation}
	D= \frac{\rho e^{\Lambda}}{\sqrt{1-v^2}}, \; S^r= \frac{\rho h v}{1-v^2}, \; \tau= \frac{\rho h }{1-v^2} - p -D.
\end{equation}   
The flux conservative equations of the relativistic hydrodynamics take then the form  
\begin{equation}
	\partial_t {\bf U} + \frac{1}{r^2}\partial_r\left(r^2 e^{\Phi-\Lambda} {\bf f}({\bf U})\right)= {\bf s}({\bf U}).
\end{equation} 
where  ${\bf U}$ is the state vector of the conserved variables, namely ${\bf U}=[D, S^{r},\tau]$. The flux vector ${\bf f}({\bf U})$  and the source vector ${\bf s}({\bf U})$ are defined by 
\begin{eqnarray}
	{\bf f}({\bf U})= &&[Dv,S^r v + p, S^r - Dv], \\
	{\bf s}({\bf U})= &&[0, (S^rv -\tau - D)e^{\Phi + \Lambda}(8\pi r p + \frac{m}{r^2}) \nn \\
	&& + e^{\Phi + \Lambda} p\frac{m}{r^2} + 2e^{\Phi-\Lambda}\frac{p}{r} ,0]. \label{HDE}
\end{eqnarray}  

The dimensionally reduced field equations, in terms of the conserved variables, for the metric functions $\Phi$ and $m$ and the evolutionary equations for the scalar field are 
\begin{widetext}
	\begin{subequations}
		\begin{align}
			&(1+\mathcal{A}) \Phi' + \mathcal{B} \frac{X^2}{r}\dot{m}
			= X^{2}\bigg( 4\pi r(s^rv+p) +\frac{m}{r^2} \bigg)  + \frac{(P^2+Q^2)r}{2} 
			+\frac{4m\lambda^2}{r^2} \bigg( \frac{d^2f}{d\varphi^2}P^2+\frac{df}{d\varphi} e^{\Lambda-\Phi}\dot{P} \bigg), \label{eq:dphidr}\\
			&(1+\mathcal{A})m'+ \mathcal{B}\dot{m}
			= 4\pi r^2(\tau+D) + \mathcal{A}\frac{m}{r} + \frac{r^2}{2X^2}(P^2+Q^2)
			+\frac{4m\lambda^2}{rX^2} \bigg( \frac{d^2f}{d\varphi^2}Q^2+\frac{df}{d\varphi} Q' \bigg), \label{eq:dmdr}\\
			&\mathcal{C} m'+ (1+\mathcal{A})\dot{m} 
			= -4\pi r^2\alpha X^{-1}s^r + \mathcal{C} \frac{m}{r}  + e^{\Phi-3\Lambda}r^2PQ + \frac{4m\lambda^2}{r}e^{\Phi-3\Lambda}
			\bigg( \frac{d^2f}{d\varphi^2}PQ+\frac{df}{d\varphi} P' \bigg), \label{eq:dmdt}\\  & \nonumber \\ & \nonumber \\
			&\partial_{t}P= e^{\Phi-\Lambda}Q' + e^{\Phi-\Lambda}(\Phi'-\Lambda'+\frac{2}{r})Q-\frac{1}{4}\frac{dV(\varphi)}{d\varphi}e^{\Phi+\Lambda} \nonumber\\
			&\qquad +2\frac{df(\varphi)}{d\varphi}\frac{\lambda^2}{r^2}e^{\Phi+\Lambda}
			\biggl\{ (1-e^{-2\Lambda})  
			\bigg[ \frac{1}{r}(\Phi'-\Lambda')e^{-2\Lambda}-8\pi p+ (Q^2-P^2)e^{-2\Lambda} +\frac{\Gamma_{\theta\theta}}{r^2} \bigg]
			-2\Phi'\Lambda'e^{-4\Lambda}+2\dot{\Lambda}^2e^{-2\Phi-2\Lambda}
			\biggr\},\label{eq:P}
		\end{align}
		
		\begin{align}
			\partial_{t}Q=& e^{\Phi-\Lambda} [P' + (\Phi'-\Lambda')P]. \label{eq:Q}
		\end{align}
	\end{subequations}
\end{widetext}
where $Q= \partial_r\varphi$ and $P=e^{\Lambda -\Phi}\partial_t\varphi$ .

We see that the scalar field couples to the metric functions via 
\begin{subequations}                         
	\begin{align}
		\mathcal{A} = \lambda^2Q\frac{df}{d\varphi}\frac{2}{r}(1-3e^{-2\Lambda}),
	\end{align}
	\begin{align}
		\mathcal{B} = \frac{4m\lambda^2}{r^2}e^{\Lambda-\Phi} P\frac{df}{d\varphi},
	\end{align}
	and
	\begin{align}
		\mathcal{C} = \frac{4m\lambda^2}{r^2}e^{\Phi-\Lambda} P\frac{df}{d\varphi}.
	\end{align}
\end{subequations}

The aforementioned equations are solved numerically in the following scheme. Equations \eqref{eq:dmdr} and \eqref{eq:dmdt} form a system of equations
\begin{align}\label{eq:massfunc}
	\begin{pmatrix}
		1+\mathcal{A} & \mathcal{B} \\
		\mathcal{C} & 1+\mathcal{A}
	\end{pmatrix}
	\begin{pmatrix}
		m'\\
		\dot{m}
	\end{pmatrix}
	=
	\begin{pmatrix}
		\text{LHS of }\eqref{eq:dmdr} \\
		\text{LHS of }\eqref{eq:dmdt}
	\end{pmatrix};
\end{align}
however, equation \eqref{eq:dphidr} should be solved for simultaneously with the $\theta\theta$ component of the equation for $\Gamma_{\mu\nu}$ (can be found in the main text of the paper) and \eqref{eq:P}, symbolized by
\begin{align}\label{eq:coupledfunc}
	\textbf{T}
	\begin{pmatrix}
		\Phi'\\
		\dot{P}\\
		\Gamma_{\theta\theta}r^{-2}
	\end{pmatrix}
	=\textbf{V}
\end{align}
where $\textbf{T}$ and $\textbf{V}$ can be obtained after a lengthy manipulation of relevant equations.

In the numerical solution of this system we follow the scheme of \cite{Ripley19c}, and solve the mass function [Eq.~\eqref{eq:massfunc}] and the redshift $\Phi$ [Eq.~\eqref{eq:coupledfunc}] iteratively until the computational error drops below a predefined threshold. In solving Eq.~\eqref{eq:coupledfunc}, $\dot{P}$ is obtained simultaneously, which then is used to evolve the scalar field.

In order to close the hydrodynamical system of equations we have to specify the equation of state (EOS) giving the pressure and other thermodynamical quantities as a function of the mass density, internal energy  and possibly the chemical composition. In the simulation, a hybrid equation EOS is used to account for the stiffening of the nuclear matter at nuclear density $\rho_{\rm nucl}=2\times 10^{14} g\, cm{^{-3}}$  and to model the response of the shocked material. The pressure consists of a cold and a thermal part, $p=p_c + p_{\rm th}$ and $e=e_c + e_{\rm th}$, where the cold parts of the pressure and the internal energy are given by 
\begin{eqnarray}
	&&p_c= K_1 \rho^{\Gamma_1}, \;\;\;  e_c= \frac{K_1}{\Gamma_1 -1 } \rho^{\Gamma_1 - 1}, \;\; {\rm for} \;\; \rho \le \rho_{\rm nucl} , \label{eq:EOS1}\\
	&&p_c= K_2 \rho^{\Gamma_2}, \;\;\;  e_c= \frac{K_2}{\Gamma_2 -1 } \rho^{\Gamma_2 - }, \;\; {\rm for} \;\; \rho > \rho_{\rm nucl} , \label{eq:EOS2}
\end{eqnarray}
and for the thermal part we assume $p_{\rm th}=(\Gamma_{\rm th} -1)\rho (e-e_c)$.

At the end let us briefly comment on the grid we employ. Due to the fact that the functions we compute exhibit much stronger spatial variation
in the central region of the star compare to the outer wave zone, the computational domain in the {\tt GR1D} code \cite{OConnor:2009iuz,Gerosa:2016fri} is divided in two parts -- an inner grid with a constant spacing and an outer grid where the spacing increases exponentially. This provides us with the opportunity to follow the dynamics in the inner core with a good resolution while allowing to extend the computational domain to large distances at acceptable computational cost. For an extensive discussion on the grid construction in the  {\tt GR1D} code we refer the reader to \cite{OConnor:2009iuz,Gerosa:2016fri}. Unless otherwise specified, in our simulations the two pieces of the grid are matched at $40$ km and the outer boundary is located at $9\times 10^5$ km. The inner grid bin (cell width) is equal to $15$m.

\section{Appendix B. Hyperbolicity}

If strong enough scalar field is realised, the evolutionary differential equations will lose locally their hyperbolicity, i.e., the discriminant of the characteristic equation $D$  becomes negative at some places \cite{Ripley19a,Ripley_2020}. As a consequence, the set of equations is no longer well-posed in the sense that the solution is not unique for the mix-type differential equations, and thus the predictability of the theory is polluted. However, for certain range of the parameters the influence of scalar field on the matter sector will maintain small enough that the hyperbolicity holds everywhere throughout the evolution \cite{Ripley19c,Ripley_2020,East:2021bqk}. In particular, for the framework adopted in this Letter, the scalar hair of an equilibrium is weaker with increasing $\beta$ when $\lambda$ is fixed \cite{Doneva_2018a}. The hyperbolicity will be preserved throughout the dynamical formation of the equilibrium for small enough $\beta$ therefore. 

\begin{figure}
	\centering
	\includegraphics[scale=0.45]{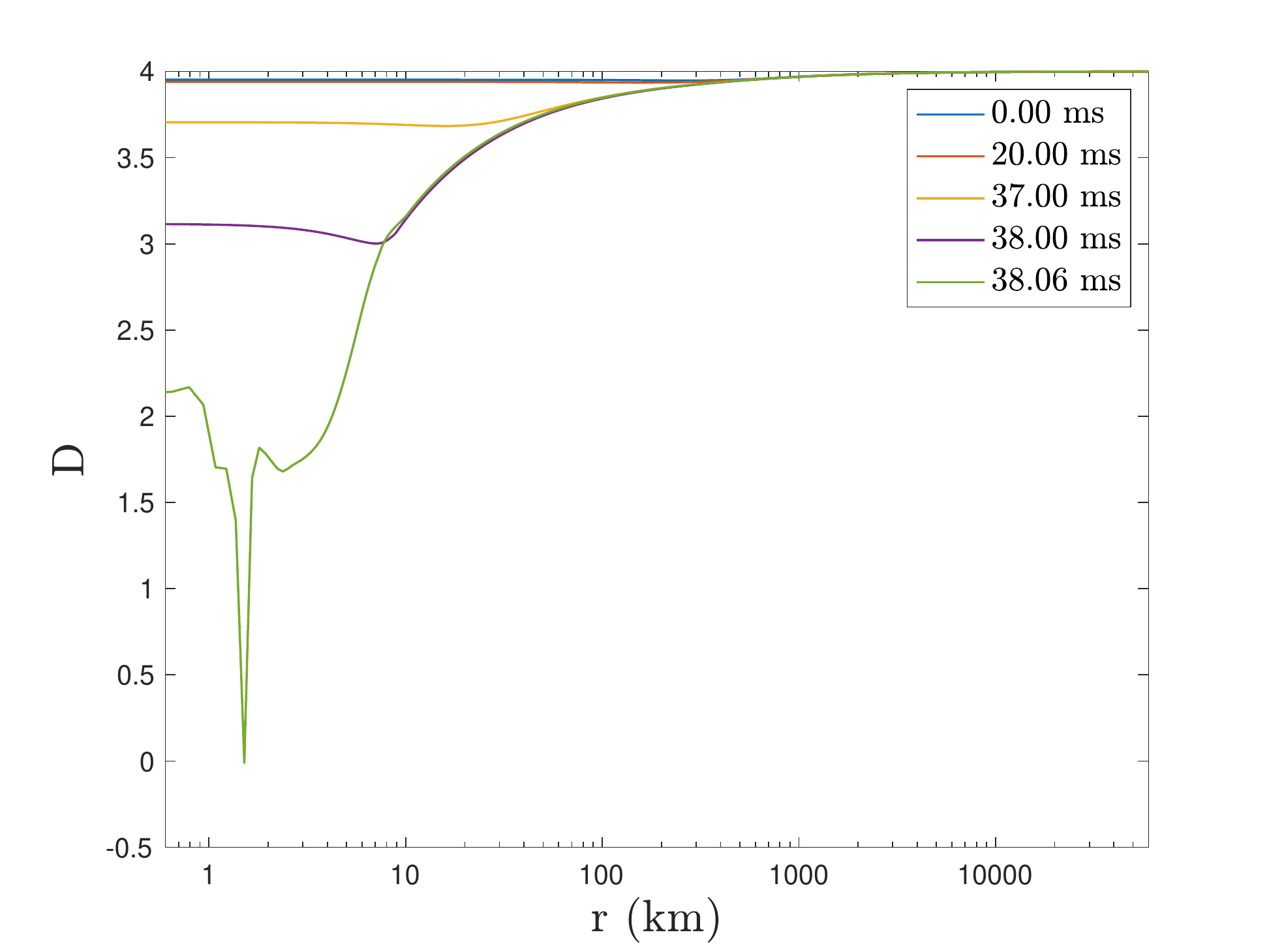}
	\caption{The discriminant of the characteristic equation as a function of $r$ at several moments. We consider z12 progenitor, and set $\lambda=80$ and $\beta=5000$. The chosen parameters and the progenitor model correspond to the channel ``$\epsilon=-1$: CC$-\varphi$NS''.}
	\label{fig:D}	
\end{figure}

For the channel ``$\epsilon=-1$: CC$-\varphi$NS'' (plotted in Fig.~3 in the main text of the Letter), we plot in Fig.~\ref{fig:D} the discriminant of the characteristic equation as a function of $r$ for $\lambda=80$ and $\beta=5000$ at several moments. The numerical evolution of the model crashes at $\sim$38.06 ms due to the violation of hyperbolicity at $\sim$1.51 km, which indicates the onset of the ill-posedness of the evolutionary equations.

\section{Appendix C. Consistency and convergence tests}
In this section we detail the reliance of our numerical results with two-fold reasoning.  The equations we utilised to evolve the metric functions $\Phi$ and $\Lambda$ are three of the four extended Einstein's field equation; the solved functions must then obey the remaining one, viz.~the $\{\theta\theta\}$-component of the extended Einstein's field equation. Denoting the difference between the right and the left hand sides of the extended Einstein's field equation as $E_{\theta\theta}(r)$, which in the perfect simulation should vanish everywhere, we measure the self-consistency of our code via the $L^2$ norm of $E_{\theta\theta}$ defined by
\begin{align}\label{eq:Ett}
	\mathcal{E} = \frac{\int(E_{\theta\theta})^2dr}{\text{grid number}},
\end{align}
where we divide the usual $L^2$-norm by the grid number or else the parasitical truncate error at each grid will pile up.

\begin{figure}
	\centering
	\includegraphics[scale=0.45]{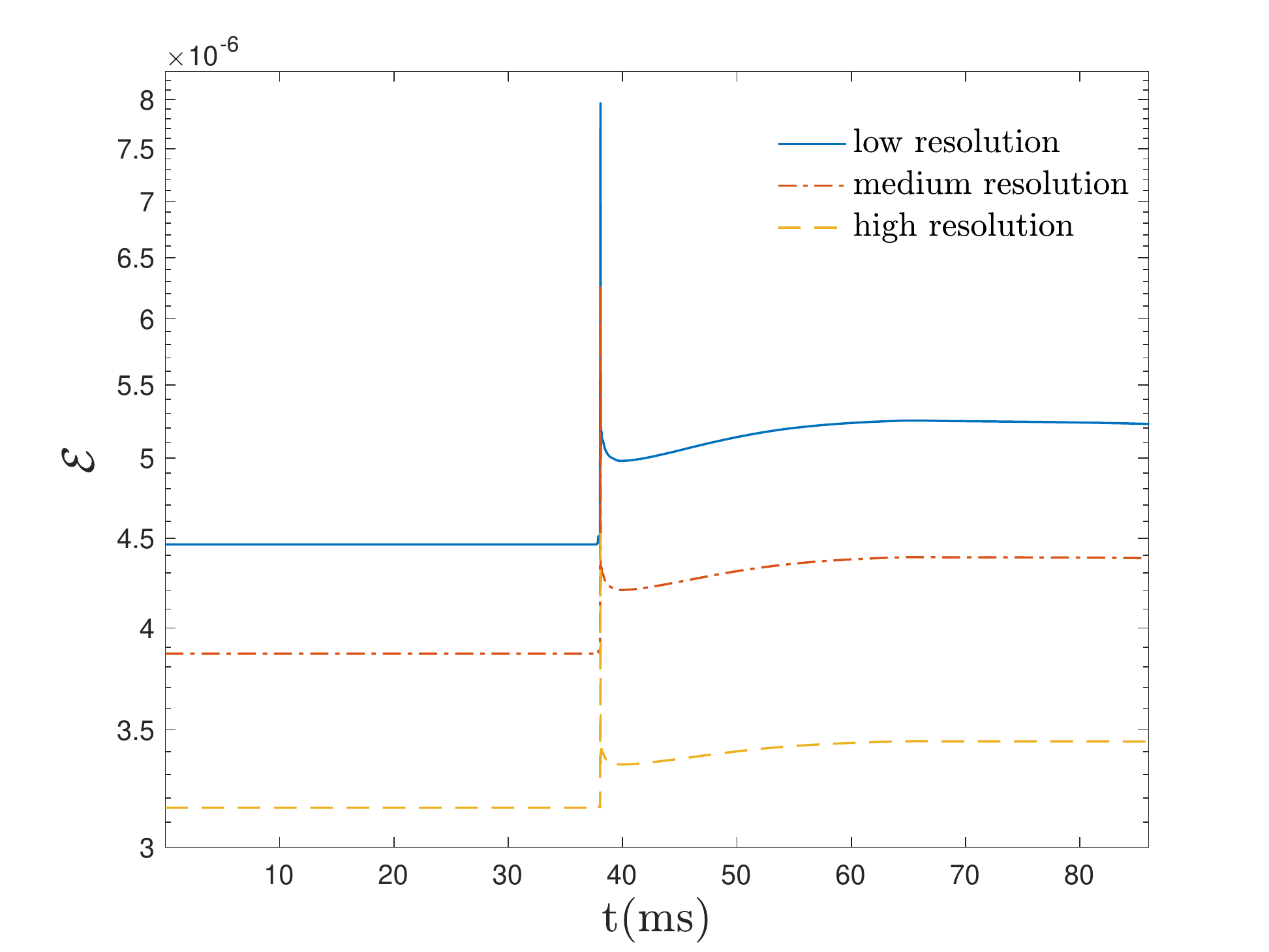}
	\caption{The $L^2$ norm of $E_{\theta\theta}$ [Eq.~\eqref{eq:Ett}] as a function of time for the case ``$\epsilon=-1$:~CC-$\varphi$NS'' discussed in the main test of the Letter with three resolutions.
	}
	\label{fig:codetest}
\end{figure}

In Fig.~\ref{fig:codetest}, we plot $\mathcal{E}$ as a function of time for the case ``$\epsilon=-1$:~CC-$\varphi$NS'' discussed in the main test of the Letter with three resolutions; in particular, the grid sizes are chosen to have the ratio of $2:1.5:1$ with the coarsest grid chosen to be the same as the one used in our simulations. We see that the residuals of the $\{\theta\theta\}$-component of the extended Einstein equations are $\lesssim 10^{-5}$ and have a sharp peak around the time when scalarization happens. In addition, the heights of this peak and the overall value of  $\mathcal{E}$  decreases with increasing grid resolution.

The convergence of the code can be quantified by comparing a certain quantity $q$ yet computed with different resolutions; stipulating we solve $q$  with three resolutions characterised by the bin sizes of the grid points $\Delta r_{\text{low}}$, $\Delta r_{\text{medi}}$, and $\Delta r_{\text{high}}$, the convergence order $n$ for $q$ is defined through
\begin{align}\label{eq:conv}
	\frac{q_{\text{low}}-q_{\text{medi}}}{q_{\text{medi}}-q_{\text{high}}}=\frac{(\Delta r_{\text{low}})^n-(\Delta r_{\text{medi}})^n}{(\Delta r_{\text{medi}})^n-(\Delta r_{\text{high}})^n},
\end{align}
where $q_{\text{low}}$, $q_{\text{medi}}$, and $q_{\text{high}}$ are the particular quantity $q$ solved with low, medium, and high resolutions. We note that Eq.~\eqref{eq:conv} is designed for uniform grid simulation.

The convergence order of the  {\tt GR1D} code is mainly influenced by the fluid dynamics.  The reason is that as shock forms and propagates outwards the convergence deteriorates from second to first order that is  characteristic feature of high-resolution shock-capturing schemes. This ultimately influences the scalar field as well since its evolution is coupled to the fluid. Thus, the convergence order will vary in time as well as location within the computational domain. For an extensive discussion and quantification of these effects we refer the reader to \cite{Gerosa:2016fri}. In order to test convergence of our modification of the  {\tt GR1D} code we followed \cite{Gerosa:2016fri} and we collapsed a polytropic star with a low central density, thus large radius, instead of z12 and z14. The reason falls in that Eq.~\eqref{eq:conv} is designed for uniform grid simulation, while we adopt for the latter progenitors a grid whose bin size is uniform inside $r=40$ km but then starts exponentially increasing until reaching to the outer boundary ($r=9\times 10^5$ km). Within this setup we managed to confirm that the convergence is of the same order as for the original  {\tt GR1D} code  \cite{Gerosa:2016fri}. Generally speaking, we see that the convergence order of our code is of second order prior to the formation of sharp wave due to the bounce, which also triggers the scalarization, while it becomes first order afterwards.

\section{Appendix D. Core-collapse scenarios}
For completeness, here we will briefly present the last two core-collapse scenarios, namely ``$\epsilon=1$: CC-$\varphi$NS'' and ``$\epsilon=1$: CC-$\varphi$NS-$\varphi$BH''. We skip of course the trivial option when the parameters are chosen in such a way that neither the neutron star nor the black hole scalarize.

\begin{figure}
	\centering
	\includegraphics[scale=0.4]{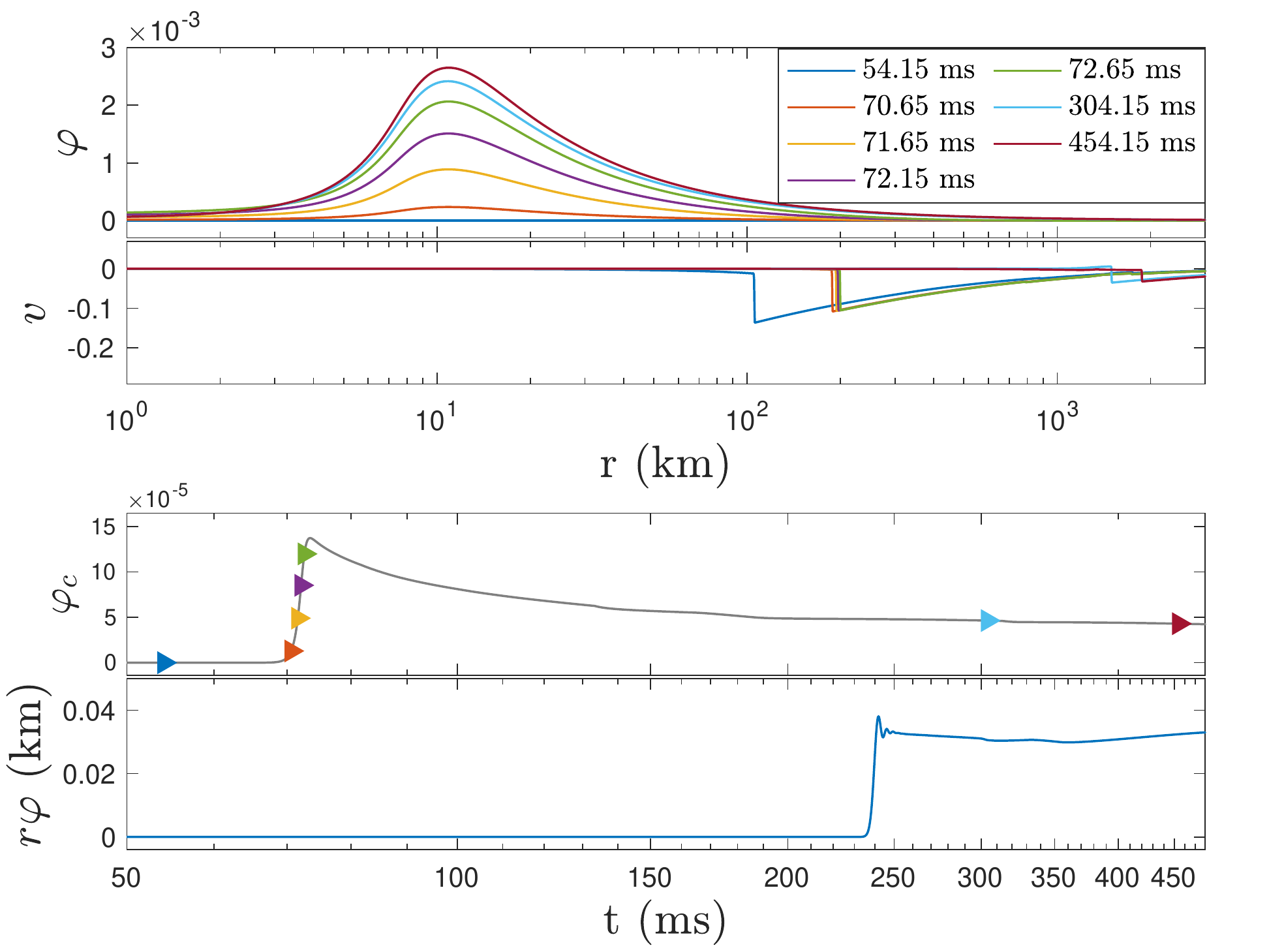}
	\caption{(collapse to a scalarized neutron star ``$\epsilon=1$: CC-$\varphi$NS'')  Temporal snapshots of scalar field $\varphi$ and fluid velocity $v$ as functions of the distance from the core $r=0$ for the z12 progenitor are plotted in the upper panel. In the lower panel the evolution of the central value of the scalar field $\varphi_c$ and the scalar charge $r\varphi$ taken at a very large distance, $50$ $000$ km, are displayed. The time delay between the formation of protoneutron star and the excitation of scalar charge is consistent with the fact that scalar fields propagate with the light speed in scalar-Gauss-Bonnet gravity. Markers in the bottom panels indicate the time of snapshots having the same colors in the corresponding upper panels. We have chosen $\epsilon=1, \lambda=100$, and $\beta=14$ $000$.}
	\label{fig:NS_remnant2}
\end{figure}

We will start with the ``$\epsilon=1$: CC-$\varphi$NS'' case that leads to the formation of a scalarized protoneutron star for the z12 progenitor, depicted in Fig. \ref{fig:NS_remnant2}.  As observed in \cite{Doneva_2018a}, for the same mass neutron star, the case with $\epsilon=1$ requires larger $\lambda$ to scalarized compared to $\epsilon=-1$ and that is why we have chosen to work with $\lambda=120$. Qualitatively the scalar field evolution has the same characteristic as for the ``$\epsilon=1$: CC-$\varphi$NS'' case. The only major difference is the scalar field profile that is drastically different for positive and negative $\epsilon$. The reason lies in the fact that $\mathcal{R}^2_{\rm GB}$  of the protoneutron star remnant is positive(negative) for $r\gtrsim(\lesssim)$ 7.77 km, implying the source term in the scalar field equation has the minimal value at some point $r<(>)7.77$ km for $\epsilon=-1(+1)$. Naturally, the scalar field will have a maximum where the source term has a negative minimum that explains the differences in the scalar field profile.

\begin{figure}
	\centering
	\includegraphics[scale=0.4]{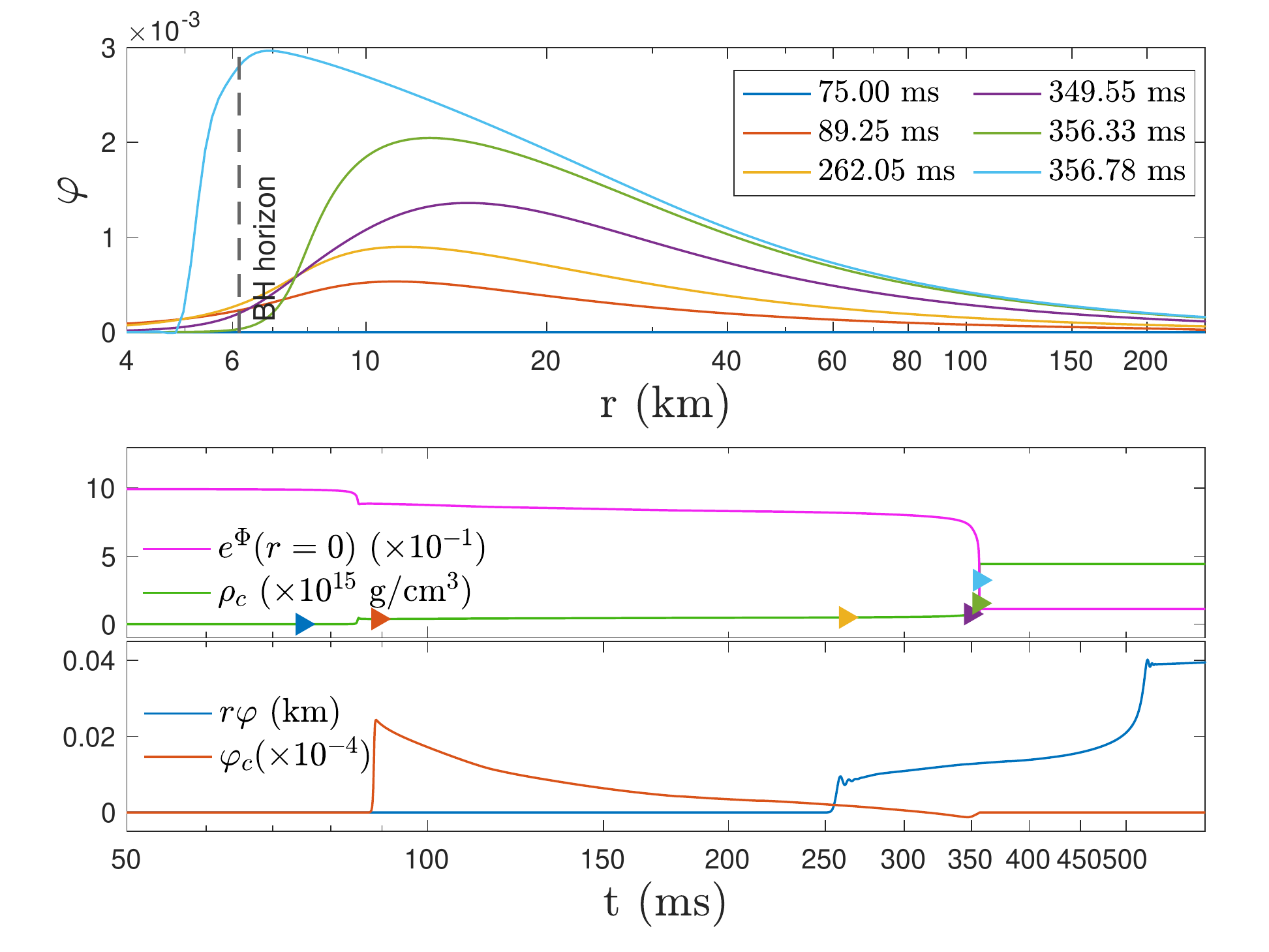}
	\caption{(collapse to a scalarized black hole through a scalarized protoneuton  star ``$\epsilon=1$: CC-$\varphi$NS-$\varphi$BH'' ) Temporal snapshots of scalar field $\varphi$ as a function of the distance from the core $r=0$ is potted for the z40 progenitor in the upper panel. In the lower group of panels central value of scalar field $\varphi_c$ and the scalar charge $r\varphi$ taken at a very large distance, $50$ $000$ km, are displayed as functions of time. We have taken $\epsilon=1$, $\lambda=120,\beta=900$ $000$ }
	\label{fig:NS_BH_remnant2}
\end{figure}

The second core-collapse scenario we will consider here is ``$\epsilon=1$: CC-$\varphi$NS-$\varphi$BH'', where a collapse to a scalarized black hole, through a scalarized protoneutron stars, is observed. This case is depicted in Fig. \ref{fig:NS_BH_remnant2}. We find that, in accordance with \cite{Doneva_2018a}, the channel for which scalarization is present in the middle state of protoneutron star (``$\epsilon=1$: CC-$\varphi$NS-$\varphi$BH'') needs larger $\lambda$ compared to ``$\epsilon=1$: CC-NS-$\varphi$BH''. The scalarization is clearly visible e.g. in the evolution of the scalar charge where the collapse to an intermediate protoneutron stars produces a nonzero scalar charge. The charge is further enhanced during the black hole formation and remains practically constant afterwards.


\begin{thebibliography}{99}
		
		\bibitem{Zwiebach:1985uq} B. Zwiebach, Phys. Lett. B  {\bf 156}, 315 (1985).  
		
		\bibitem{Gross:1986mw} D. Gross and J. H. Sloan, Nucl. Phys. B  {\textbf 291}, 41 (1987).
		
		\bibitem{Doneva_2018} D. D. Doneva and S. S. Yazadjiev, Phys. Rev. Lett. {\bf 120}, 131103 (2018).
		
		\bibitem{Silva_2018} H. Silva, J. Sakstein, L. Gualtieri, T. Sotiriou and E. Berti, Phys. Rev. Lett. {\bf 120}, 131104 (2018).
		
		\bibitem{Antoniou:2017acq}
		G.~Antoniou, A.~Bakopoulos and P.~Kanti,
		Phys. Rev. Lett. \textbf{120}, 131102 (2018).
		
		\bibitem{Doneva_2018a} D. D. Doneva and S. S. Yazadjiev, 
		J. Cosmol. Astropart. Phys. 04, 011 (2018).
		
		\bibitem{Doneva21}
		D.~D.~Doneva and S.~S.~Yazadjiev,
		arXiv:2107.01738.
		
		\bibitem{Dima:2020yac}
		A.~Dima, E.~Barausse, N.~Franchini and T.~P.~Sotiriou,
		Phys. Rev. Lett. \textbf{125}, 231101 (2020).
		
		\bibitem{Doneva:2020nbb}
		D.~D.~Doneva, L.~G.~Collodel, C.~J.~Kr\"uger and S.~S.~Yazadjiev,
		Phys. Rev. D \textbf{102}, 104027 (2020).
		
		\bibitem{Doneva_2020a} D. Doneva, L. Collodel, Ch. Kruger, S. Yazadjiev,  Eur. Phys. Jour. C {\bf 80},  1205 (2020).
		
		\bibitem{Cunha_2019} P. V. Cunha, C. A. Herdeiro, and E. Radu, Phys. Rev. Lett. {\bf 123}, 011101 (2019).
		
		\bibitem{Collodel:2019kkx} L.~G.~Collodel, B.~Kleihaus, J.~Kunz and E.~Berti, Class. Quant. Grav. \textbf{37}, 075018 (2020).
		
		\bibitem{Herdeiro:2020wei}
		C.~A.~R.~Herdeiro, E.~Radu, H.~O.~Silva, T.~P.~Sotiriou and N.~Yunes,
		Phys. Rev. Lett. \textbf{126}, 011103 (2021).
		
		\bibitem{Berti:2020kgk}
		E.~Berti, L.~G.~Collodel, B.~Kleihaus and J.~Kunz,
		Phys. Rev. Lett. \textbf{126}, 011104 (2021).
		
		

		\bibitem{Doneva:2021dqn}
		D.~D.~Doneva and S.~S.~Yazadjiev,
		Phys. Rev. D \textbf{103}, 064024 (2021).
		
		\bibitem{Ripley_2020} J. L. Ripley and F. Pretorius, Classical Quantum Gravity {\bf 37}, 155003 (2020).


		\bibitem{East:2021bqk}
		W.~E.~East and J.~L.~Ripley,
		Phys. Rev. Lett. \textbf{127}, 101102 (2021).

		
		\bibitem{Silva:2020omi}
		H.~O.~Silva, H.~Witek, M.~Elley and N.~Yunes,
		Phys. Rev. Lett. \textbf{127}, 031101 (2021).
				
		\bibitem{Benkel:2016kcq}
		R.~Benkel, T.~P.~Sotiriou and H.~Witek,
		Phys. Rev. D \textbf{94}, 121503(R) (2016).

		
		\bibitem{Novak:1999jg}
		J.~Novak and J.~M.~Ibanez,
		Astrophys. J. \textbf{533}, 392 (2000).
		
		\bibitem{Gerosa:2016fri}
		D.~Gerosa, U.~Sperhake and C.~D.~Ott,
		Class. Quant. Grav. \textbf{33}, 135002 (2016).
		
		\bibitem{Sperhake:2017itk}
		U.~Sperhake, C.~J.~Moore, R.~Rosca, M.~Agathos, D.~Gerosa and C.~D.~Ott,
		Phys. Rev. Lett. \textbf{119}, 201103 (2017).
		
		\bibitem{Cheong:2018gzn}
		P.~C.~K.~Cheong and T.~G.~F.~Li,
		Phys. Rev. D \textbf{100}, 024027 (2019).
		
		\bibitem{Rosca-Mead:2019seq}
		R.~Rosca-Mead, C.~J.~Moore, M.~Agathos and U.~Sperhake,
		Class. Quant. Grav. \textbf{36}, 134003 (2019).
		
		\bibitem{Geng:2020slq}
		C.~Q.~Geng, H.~J.~Kuan and L.~W.~Luo,
		Eur. Phys. J. C \textbf{80}, 780 (2020).
		
		\bibitem{Rosca-Mead:2020ehn}
		R.~Rosca-Mead, U.~Sperhake, C.~J.~Moore, M.~Agathos, D.~Gerosa and C.~D.~Ott,
		Phys. Rev. D \textbf{102}, 044010 (2020).
		
		\bibitem{Stefanov:2007eq}
		I.~Z.~Stefanov, S.~S.~Yazadjiev and M.~D.~Todorov,
		Mod. Phys. Lett. A \textbf{23}, 2915 (2008).
		
		\bibitem{Doneva:2010ke}
		D.~D.~Doneva, S.~S.~Yazadjiev, K.~D.~Kokkotas and I.~Z.~Stefanov,
		Phys. Rev. D \textbf{82}, 064030 (2010).
		
		\bibitem{Cardoso:2013fwa}
		V.~Cardoso, I.~P.~Carucci, P.~Pani and T.~P.~Sotiriou,
		Phys. Rev. Lett. \textbf{111}, 111101 (2013).
		
		\bibitem{OConnor:2009iuz}
		E.~O'Connor and C.~D.~Ott,
		Class. Quant. Grav. \textbf{27}, 114103 (2010).
		
		
		\bibitem{Ripley19c}
		J.~L.~Ripley and F.~Pretorius,
		Phys. Rev. D \textbf{101}, 044015 (2020).
		
		\bibitem{Doneva:2018rou}
		D.~D.~Doneva, S.~Kiorpelidi, P.~G.~Nedkova, E.~Papantonopoulos and S.~S.~Yazadjiev,
		Phys. Rev. D \textbf{98}, 104056 (2018).
		
		
		\bibitem{Popov:2012ng}
		S.~B.~Popov and R.~Turolla,
		Astrophys. Space Sci. \textbf{341}, 457 (2012).
		
		\bibitem{Noutsos:2013ce}
		A.~Noutsos, D.~Schnitzeler, E.~Keane, M.~Kramer and S.~Johnston,
		Mon. Not. Roy. Astron. Soc. \textbf{430}, 2281 (2013).
		
		\bibitem{Woosley:2007as}
		S.~E.~Woosley and A.~Heger,
		Phys. Rept. \textbf{442}, 269 (2007).
		
		\bibitem{Dimmelmeier07}
		H.~Dimmelmeier, C.~D.~Ott, H.~T.~Janka, A.~Marek and E.~Mueller,
		Phys. Rev. Lett. \textbf{98}, 251101 (2007).
		
		\bibitem{Dimmelmeier08}
		H.~Dimmelmeier, C.~D.~Ott, A.~Marek and H.~T.~Janka,
		Phys. Rev. D \textbf{78}, 064056 (2008).
		
		\bibitem{Shen11}
		H.~Shen, H.~Toki, K.~Oyamatsu and K.~Sumiyoshi,
		Astrophys. J. Suppl. \textbf{197}, 20 (2011).
		
		\bibitem{LS}
		J.~M.~Lattimer and F.~D.~Swesty,
		Nucl. Phys. A \textbf{535}, 331 (1991).
		
		\bibitem{Ripley19a}
		J.~L.~Ripley and F.~Pretorius,
		Phys. Rev. D \textbf{99}, 084014 (2019).
		
		\bibitem{Radice18}
		D.~Radice, V.~Morozova, A.~Burrows, D.~Vartanyan and H.~Nagakura,
		Astrophys. J. Lett. \textbf{876}, L9 (2019).

		
		\bibitem{Abdikamalov:2020jzn}
		E.~Abdikamalov, G.~Pagliaroli and D.~Radice,
		arXiv:2010.04356.

		
	\end{thebibliography}
\end{document}